\documentstyle[12pt]{article}

\newcommand{\be}{\begin{equation}}
\newcommand{\ee}{\end{equation}}
\newcommand{\ba}{\begin{eqnarray}}
\newcommand{\ea}{\end{eqnarray}}

\newcommand{\tr}{\rm tr}

\begin{document}
%\hoffset=-.4truein\voffset=-0.5truein
%\setlength{\baselineskip}{14pt}
\setlength{\textheight}{8.5 in}
\begin{titlepage}
\begin{center}
\hfill LPTENS 08-53\\
\vskip 0.6 in
{\large \bf{ Computing topological invariants with one and two-matrix models}}
\vskip .6 in
\begin{center}
{\bf E. Br\'ezin$^{a)}$}{\it and} {\bf S. Hikami$^{b)}$}
\end{center}
\vskip 5mm
\begin{center}
{$^{a)}$ Laboratoire de Physique
Th\'eorique, Ecole Normale Sup\'erieure}\\ {24 rue Lhomond 75231, Paris
Cedex
05, France. e-mail: brezin@lpt.ens.fr{\footnote{\it
Unit\'e Mixte de Recherche 8549 du Centre National de la
Recherche Scientifique et de l'\'Ecole Normale Sup\'erieure.
} }}\\
{$^{b)}$ Department of Basic Sciences,
} {University of Tokyo,
Meguro-ku, Komaba, Tokyo 153, Japan. e-mail:hikami@dice.c.u-tokyo.ac.jp}\\
\end{center}     
\vskip 3mm         

{\bf Abstract}                  
\end{center}

A generalization of the Kontsevich Airy-model allows one to compute the intersection numbers of the moduli space of $p$-spin curves. These models are deduced from averages of characteristic polynomials over 
Gaussian ensembles of random matrices in an external matrix source. 
After use of a duality, and of an appropriate tuning of the source,  we obtain in a double scaling limit these intersection numbers as polynomials in $p$.
One can then take the   limit  $p\to -1$ which yields a matrix model for  orbifold Euler characteristics.
The generalization to a time-dependent matrix model, which is equivalent to a  
two-matrix model, may be treated along the same lines  ;  
it also yields a logarithmic potential with additional vertices for general $p$.

\end{titlepage}
\vskip 3mm

%*******************************
\section{Introduction}

Many topological invariants 
have been  computed from matrix models of moduli spaces. The well-known Kontsevich's 
Airy matrix model \cite{Kontsevich} gives  the intersection numbers of  
the moduli spaces of curves, which was also studied by the double scaling limit of one
matrix model \cite{Ambjorn} . 
The  Euler characteristics  of  orbifolds have been computed from  
the Penner model \cite{Penner}.
The intersection numbers for the $p$-spin curves are obtained from 
the generalized Kontsevich model \cite{Witten1}.
These matrix models all provide 
explicit results for the intersection numbers.

In this article, we  discuss  the above models, which lead to three kinds of  matrix models, in 
a unified way.
Our  formulation starts from  simple Gaussian
matrix models with an external matrix source. 
In recent articles we have already considered the 
average characteristic polynomials in these Gaussian ensembles, 
and derived, through a duality 
relation and the replica method, 
the intersection numbers of $p$-spin curves  \cite{BH1,BH2,BH3}. 
This duality relates the average of the  product of $k$ characteristic polynomials 
for $N\times N$ random matrices $M$,  
to the average of the product of $N$ characteristic polynomials over $k\times k$  Gaussian  
random matrices $B$. 
In the large N limit, the matrix model for $B$ reduces to
the higher Airy matrix \cite{BH2,BH3} for the intersection numbers of spin 
curves studied by Witten\cite{Witten1,Witten}. 

This duality  allows one to compute
the intersection numbers for the spin moduli spaces with 
$n$-marked points and genus $g$,  
from an  $n$-point correlation function $U(s_1,...,s_n)$
of Gaussian random matrices in a scaling limit near critical edges \cite{BH8,BH9}. 
The basic steps are recalled in section 2.

In this article we first compute explicitly 
the intersection numbers of moduli space of $p$-spin curves
with one marked point, for arbitrary values of $p$, as  polynomials in $p$.
We have obtained earlier the intersection numbers for $p$=2,3 and 4 explicitly, but we discuss here  arbitrary $p$.
This allows us to consider continuations in $p$ ; in particular 
the limit $p\to -1$  exhibits an interesting relation between the intersection numbers,
($\tau$-class) and the orbifold Euler characteristics $\chi(M_{g,1})=
\zeta( 1- 2g)$  ($\zeta(x)$ is Riemann zeta function) \cite{Penner,HZ}. 
In section 3 we derive these numbers for surfaces with one marked point. 

 In section 4 we show that the intersection numbers with n-marked points 
for p-spin curves are  also obtained easily
from the integral representation of $U(s_1,...,s_n)$ at  
the critical values tuned through an appropriate external source.
We evaluate the case of two marked points for genus one (g=1) and any $p$. The generating function
$U$ is given for three and four marked points in Appendices A and B. Our results through
this generating function $U$ are consistent with the previous recursive results of
Virasoro equations. We find that
the ring structure of the primary fields, for genus zero, 
is deduced from these n-point correlation functions $U$
for arbitrary $p$. This shows   
that the random matrix theory with external source near critical edges,
has a structure of a minimal $N=2$  superconformal field theory with Lie algebra $A_{p-1}$ .

It has been conjectured by Witten  that the free energy $F$ which 
generates the intersection numbers of the moduli space of $p$-spin curves satisfies
a Gelfand-Dikii hierarchy \cite{Witten1,Gelfand}. We show here that the intersection numbers, 
computed  from the integral representation
of $U(s_1,...,s_n)$, do satisfy Gelfand-Dikii equations.
We present in Appendix C, this Gelfand-Dikii hierarchy
equations and the
construction of the super potential for the primary fields.
We also note that, with respect to Witten's conjecture, 
that the definition of  intersection numbers as
an integral over the compactified moduli space $\bar M_{g,n}$, is 
similar in structure to the integral representation of $U(s_1,...,s_n)$. 

In section 5, begins a second part of the paper devoted to the 
the time dependent  Gaussian matrix model 
for which we extend our previous work on duality   \cite{BH1,BH2,BH3}.
The time-dependent model, a matrix quantum mechanics of harmonic oscillators, 
reduces easily, for a Gaussian distribution, to
an equivalent two-matrix model. Again one may derive (section 6)  a 
dual model in the presence of a matrix source. 
We then obtain, with an appropriate tuning of the source, 
the two-matrix equivalents of the  Kontsevich plus Penner models 
for matter with central charge
$c=1$ for the $p=2$ case. For $p > 2$, additional terms
are present with respect to the c=1 matrix model for 
tachyon correlators \cite{Mukhi1,Dijkgraaf2}.

\section{Replica and duality for the one-matrix model}

Let us first summarize   the steps which one uses to obtain Airy 
and higher Airy matrix models from the Gaussian one-matrix model in a source, 
followed by the computation of  the intersection numbers through the replica method
\cite{BH1,BH2,BH3}.

The m-point correlation functions of the eigenvalues in the Gaussian unitary ensemble
are conveniently deduced from their Fourier transforms $U(s_1,...,s_m)$,  defined as
\ba\label{U1m}
U(s_1,s_2,...,s_m) &=& < {\tr} e^{s_1 M} {\tr}e^{s_2 M} \cdots {\tr}e^{s_m M}>\nonumber\\
&=& \int \prod_{l=1}^m d \lambda_l e^{\sum i t_l \lambda_l} 
<\prod_1^m  {\tr } \delta (\lambda_j - M) >
\ea
where $s_l= i t_l$ ;  $M$ is an $N\times N$ Hermitian random matrix.
The bracket stands for averages with the Gaussian probability measure 
\be
 < X > _A= \int dM e^{-\frac{N}{2}{\tr} M^2 +N {\tr }M A} X(M),
\ee
$A$ is an $N\times N$ external Hermitian  source matrix. 
We may assume that this matrix $A$ is diagonal
with eigenvalues  $a_j$. We consider later the external source $A$ with $(p-1)$ distinct eigenvalues,
each of them being $\frac{N}{p-1}$ times degenerate. This parameter $p$ is crucial
in this paper.

Let us consider first the  one point case (m=1), namely $U(s)$. 
The replica limit $k\rightarrow 0$ for $<{\tr} \delta(\lambda - M)>$ relies on the identity
\ba
&&\lim\limits_{k\to 0}  \frac{1}{k}([{\rm det}(\lambda\cdot {\rm I} - M)]^k - 1)\nonumber\\
&&= {\tr} {\rm log}(\lambda\cdot {\rm I} - M)
\ea
Taking a derivative with respect to $\lambda$, it yields the density of states  ${\tr}\delta(\lambda - M)$  
as the imaginary part of the resolvent when 
the imaginary part of $\lambda$ goes to zero. Thus
$U(s)$ is expressed in terms of  
products of characteristic polynomials,
\be
U(s) = \lim\limits_{k\to 0} \frac{1}{k} \int d\lambda e^{s \lambda} \frac{\partial}{\partial \lambda}
< \prod_{\alpha=1}^k {\rm det}(\lambda_\alpha - M) >_A\vert _{\lambda_\alpha=\lambda}
\ee
We have  introduced a replica symmetry breaking by taking $k$ distinct
$\lambda_\alpha$ ($\alpha=1,...,k$) \cite{BH4},
in order to use the duality formula \cite{BH2,BH7},
\be\label{dualityform}
\frac{1}{Z_0}<\prod_{\alpha=1}^k {\rm det}(\lambda_\alpha - M) >_A = \frac{1}{Z'_0}< \prod_{j=1}^N
{\rm det}(a_j - i B)>_{\Lambda}.
\ee
where $\Lambda= diag(\lambda_1,...,\lambda_n)$, $B$ is an $k\times k$ Hermitian random matrix, and $Z_0$ and $Z'_0$
are normalization constants. The probability distribution for $B$ in the right hand side is
\be
<Y>_{\Lambda} = \int dB e^{-\frac{N}{2}{\tr} B^2 + i N{\tr} B \Lambda} Y(B)
\ee
The simplest case  consists of taking an external source multiple
 of the identity, namely $a_j = 1,\ j=1,\cdots N$. The effect of this source 
is simply to shift 
all the eigenvalues of $M$ by one : the left  edge of Wigner's semi-circle is now at 
the origin. In the large N limit, after exponentiation, expansion of the integrand 
in powers of B, cancellation of the ${\tr B^2}$ terms, we obtain
\be\label{Kontsevich0}
Z=\lim_{N\to \infty}<[{\rm det}(1 - i B)]^N >_{\Lambda} = \int dB e^{\frac{iN}{3}{\tr} B^3 + iN {\tr}B (\Lambda-1)}
\ee
which is Kontsevich Airy matrix-model. (We explore here a "double scaling limit", namely the vicinity of the origin, in which  the $(\lambda_a -1)$ are of order $N^{-2/3}$, $B$ of order $N^{-1/3}$  ; in this regime  $N {\tr} B^l$ is negligible for $l\geq 4$). This Airy matrix model has an expansion in terms of the moduli parameters
$t_m$,
\be\label{tm}
t_m = C {\rm tr}\frac{1}{\Lambda^{2m+1}}= C \sum_{\alpha=1}^k \frac{1}{\lambda_\alpha^{2m+1}}
\ee
\be\label{Kontsevich}
Z = \sum_{m,k_m} <\prod_{m} \tau_m^{k_m}> \prod_m \frac{t_m^{k_m}}{k_m!}
\ee
where $C$ is a normalization constant, to be determined later.
When we set the  k distinct  $\lambda_\alpha$ to a common value $ \lambda$, we have $t_m= C \frac{k}{\lambda^{2m+1}}$. Then, in the limit $k\rightarrow
0$, only single $\tau$'s appear, since the parameter $t_m$ is proportional to $k$. 
Thus the zero-replica limit $k\to 0$ yields the intersection numbers with
 one-marked point (one $t_m$) :

\be
U(s) = \int d\lambda e^{s \lambda} \frac{\partial}{\partial \lambda}[ 1 + <\tau_1> t_1 + <\tau_4> t_4 + \cdots]
\ee
where $<\tau_1>= \frac{1}{24}$, $<\tau_4> = \frac{1}{(24)^2 2!}$. 
As found in \cite{BH1}, $U(s)$ is obtained, after approriate normalization,  as
\be
U(s) = \frac{1}{s^{3/2}} e^{\frac{s^3}{24}}
\ee
which gives the intersection number $<\tau_m>$ for the moduli space of curves  
with one marked point,
\be\label{taum}
<\tau_m> = \frac{1}{(24)^g g!}
\ee
where $g$ is the genus of the curve and $m= 3g -2$.
We have thus shown that the Fourier transform $U(s)$ gives 
the intersection numbers $<\tau_m>$ of the moduli space of curves with one marked point
\cite{BH1,Okounkov}.

The replica limit $k\to 0$ ,
where the matrix $B$ is $k\times k$, was studied in \cite{BH2}, and 
it gives the intersection numbers of (\ref{taum}). Note that in the original Kontsevich model of (\ref{Kontsevich0}),
the matrix size $k$ was arbitrary, and the universal intersection numbers $<\tau_m>$ are independent of $k$.

From other  tunings of the external source $a_j$, we may obtain also 
the intersection numbers of the moduli space of $p$-spin curves \cite{Witten1}
with one marked point, which exhibit "spin structures".
Indeed we may tune the external source so that 
the asymptotic density of states vanishes at an edge as 
$\rho(\lambda) \sim \lambda^{\frac{1}{p}}$. 
This will yield the exact values for p-spin curves with genus $g$ and  
one marked point. A spin index $j=0,1,...,p-1$ is now needed. 
From this tuning of the external source, we obtain the generalized Kontsevich model,
\be\label{Kontsevich2}
Z = \int dB e^{\frac{i}{p+1}{\tr} B^{p+1} - i \tr B \Lambda^p}
\ee
where $B$ is $k\times k$. The derivation of this partition function from the right hand side 
of the duality formula (\ref{dualityform}) will be given
in the next section. This $Z$ has an expansion,
\be\label{Kontsevich3}
Z = \sum_{m,j,k_{m,j}} <\prod_{m,j} \tau_{m,j}^{k_{m,j}}> \prod_m \frac{t_{m,j}^{k_{m,j}}}{k_{m,j}!}
\ee
where
\be\label{coefficientt}
t_{m,j} = (-p)^{\frac{j-p-m(p+2)}{2(p+1)}}
\prod_{l=0}^{m-1}
(1 + j + l p) {\tr}\frac{1}{\Lambda^{p m + j+1}}.
\ee
The normalization constant $C$ in (\ref{tm}) is fixed by (\ref{coefficientt}).
The intersection numbers of moduli space of $p$-spin curves are 
defined by the integral formula of compactified
the moduli space $\bar M_{g,n}$ \cite{Witten}
\be\label{Uj}
< \tau_{n_1}(U_{j_1}) \cdots \tau_{n_s}(U_{j_s}) > = 
\frac{1}{(\hat k+2)^g}\int_{\bar M_{g,s}} C_T(\nu) \prod_{i=1}^s
(c_1 ({\mathcal L}_i))^{n_i}
\ee
where $U_j$ is an operator for the primary matter field (tachyon), related to 
top Chern class $C_T(\nu)$, and 
$\tau_{n}$ is a gravitational operator,
related to the first Chern class $c_1$ of the line bundle ${\mathcal L}_i$ 
at the $i$th-marked point.
We denote $\tau_{n}(U_j)$ by $\tau_{n,j}$, and $j$ represents the spin index (j=0,...,p-1). 
The indices $n_i$ and $j_i$ are related to genus $g$ and numbers of marked points $s$ through
\be
(p+1)(2 g - 2 + s) = \sum_{i=1}^s (p n_i + j_i + 1).
\ee

This intersection theory for spin-curves \cite{Witten1} is known to be related to the minimal $N=2$ superconformal field theory of Lie algebra $A_{p-1}$ type,
which is equivalent to $SU(2)_{\hat k}/U(1)$ Wess-Zumino-Witten model. $\hat k$ is a number of levels 
and it is related to $p$ by  $\hat k= 
p-2$.
This relation is derived from a super potential $W$ for the chiral ring structures of primary fields ;
 we will obtain this chiral structure later by the consideration 
of the n-point correlation functions $U(s_1,...,s_n)$ (see appendix C).

As remarked by Witten \cite{Witten1}, the limit $p\to -1$ ($\hat k\to
-3$) corresponds to the top Chern class without gravity decendants
$c_1({\mathcal L}_i)$, and this top Chern class becomes the orbifold Euler characteristic 
class \cite{HZ,Penner}.

For $3\le p$, we have to consider the above spin structures for the intersection numbers. 
We find the intersection number with one marked point for arbitrary genus, $<\tau_{n,j}>$ \cite{BH3} for p=3 as
\be
<\tau_{n,j}>_g = \frac{1}{(12)^g g!}\frac{\Gamma(\frac{g+1}{3})}{\Gamma(\frac{2-j}{3})}
\ee
where $n= (8g-5-j)/3$ and $j=0$ for $g=3m+1$, $j=1$ for $g=3m$ $(m=1,2,...)$.

In the replica limit, $k\rightarrow 0$ for the matrix $B$, a closed expression for  $U_0(s_1,...,s_k)$ is known  \cite{BH2}
(the surfix "0" refers to zero external source in the Gaussian probability distribution),
\be\label{replica}
\lim\limits_{k\rightarrow 0} U_0(s_1,...,s_n) 
= \frac{1}{\sigma^2} \prod_{i=1}^n 2 {\rm sinh}(\frac{s_i \sigma}{2})
\ee
where $\sigma = \sum_{i=1}^n s_i$, in which  $U_0(s_1,...,s_n)$ is defined by 

\ba\label{replica1}
U_0(s_1,...,s_n) &=& <{\tr} e^{s_1 B} \cdots {\tr} e^{s_n B} >\nonumber\\
&=& \sum_{l_i} <\prod_{i=1}^n {\tr} B^{l_i}> \prod_{i=1}^n \frac{s_i^{l_i}}{l_i!}.
\ea

From (\ref{replica}), we obtain the intersection numbers of $p$-spin curves 
with one marked point. The replica limit
$k\rightarrow 0$ selects only the one marked point ribbon graphs 
on a  genus g-Riemann surface. This method gives the  intersection numbers of 
$p$-spin curves with one marked point, and
the results 
coincide with those obtained from $U(s)$ \cite{BH2}.

For two marked points, one deals with the dual quantity  
$U(s_1,s_2)$, again at a critical edge point.
The correspondence is the same as for the one-marked point. We have
\ba\label{twopointU}
&&U(s_1,s_2) = < {\tr} e^{s_1 M} {\tr} e^{s_2 M} >\nonumber\\
&=& \int d\lambda_1 d \lambda_2 e^{s_1 \lambda_1+s_2 \lambda_2}
< {\tr} \delta(\lambda_1 - M) {\tr}\delta (\lambda_2 - M) >\nonumber\\
&=& \lim\limits_{k_1,k_2\rightarrow 0} \int d \lambda_1 
d\lambda_2 e^{s_1 \lambda_1+s_2 \lambda_2}
\frac{\partial^2}{\partial \lambda_1 \partial
\lambda_2} < [{\rm det} (\lambda_1 - M) ]^{k_1} [{\rm det} (\lambda_2 - M)]^{k_2} >\nonumber\\
&=& \lim\limits_{k_1,k_2\rightarrow 0} \int d \lambda_1 d\lambda_2 
e^{s_1 \lambda_1+s_2 \lambda_2}
\frac{\partial^2}{\partial \lambda_1 \partial
\lambda_2} <[{\rm det}(1 - i B)]^N >_{\Lambda}
\ea
where $\Lambda= diag(\lambda_1,\cdots,\lambda_1,\lambda,\cdots,\lambda_2)$, $\lambda_1$ and $\lambda_2$ are degenerate
$k_1$ and $k_2$ times respectively. The matrix $B$ is an $(k_1+k_2)\times (k_1+k_2)$ Hermitian matrix.
In the limit of zero replica, one selects the terms of order  $k_1 k_2$ in the Airy matrix model,  for instance a term like
$\frac{k_1}{\lambda_1^3} \frac{k_2}{\lambda_2^3}$, and we obtain the
two marked points contribution for the intersection numbers. The Fourier transform $U(s_1,s_2)$, with respect to $\lambda_1$
and $\lambda_2$, gives the intersection numbers as coefficients of the Taylor expansion in $s_1$ and $s_2$.
For the case  p=2, this was checked for arbitrary genus \cite{BH1} and it does yield the known values .

For higher marked points, the argument is similar.
For  $n$ marked points, one considers the terms 
\be
\sum < \tau_{m_1} \cdots \tau_{m_n}> \frac{k_1 \cdots k_n}{\lambda_1^{2 m_1+1} 
\lambda_2^{2 m_2 +1} 
\cdots \lambda_n^{2m_n+1}}
\ee
emerging from
the $B$-matrix integral. The Fourier transform of these quantities is given by 
$U(s_1,s_2,...,s_n)$.

Thus we have indeed a   method for computing
the intersection numbers of the moduli of  curves from random matrix theory, 
based on the expression for $U(s_1,...,s_n)$.

An exact and  useful integral representation for $U(s_1,...,s_n)$ is  known 
in the presence of an external 
matrix source $A$ \cite{BH5}.
\ba\label{U(sn)} 
&&U(s_1,\cdots, s_n) = \frac{1}{N} \langle \rm{tr} e^{s_1 M} \cdots \rm{tr} e^{s_n M} 
\rangle  \nonumber\\
&&= e^{\sum_1^n s_i^2}\oint \prod_1^n \frac{du_i}{2\pi i} 
e^{\sum_1^n u_is_i} \prod_{\alpha=1}^N \prod_{i=1}^n (1-\frac{s_i}{a_\alpha- u_i}) 
\det \frac{1}{u_i-u_j + s_i}
\ea
We will use this formula in the following sections to obtain the intersection numbers of $p$-spin curves of
arbitrary genus $g$ for n-marked points, through an appropriate  tuning of the
external source $a_j$, in a scaling  large N limit.

\section{ The $p$-dependence of the intersection numbers with one marked point}

The partition function $Z_p$ for the generalized Kontsevich model 
is given by the $k\times k$ Hermitian matrix $B$,
\be\label{generalized}
Z_p= \int dB e^{\frac{1}{p+1} {\tr} B^{p+1} - {\tr} B \Lambda}
\ee
This model is obtained, after use of the duality,  from the expectation values of characteristic
polynomials  (\ref{dualityform}). We take here an
external source $A$ with $(p-1)$ distinct eigenvalues, each of them being  $\frac{N}{p-1}$ 
times degenerate :   $A= {\rm diag}(a_1,...,a_1,....,a_{p=1},....,a_{p-1})$.
After duality, the expectation values of characteristic polynomials become
\be
<\prod_{\alpha=1}^{p-1} {\rm det}(a_\alpha - i B)^{\frac{N}{p-1}}>=< {\rm exp}[ \sum_{\alpha=1}^{p-1}
{\tr} {\rm log}(1 - \frac{i B}{a_\alpha}) + N \sum {\rm log}(\prod_{\alpha=1}^{p-1} a_\alpha)]>
\ee
We now specify the $(p-1)$ distinct eigenvalues of the external source by  the $(p-1)$  conditions :
\ba\label{acond}
&&\sum_{\alpha}^{p-1} \frac{1}{a_\alpha^2} = p-1, \hskip 5mm 
\sum_{\alpha=1}^{p-1} \frac{1}{a_\alpha^m} = 0, \hskip 5mm (m=3,4,...,p)\nonumber\\
&&\sum_{\alpha=1}^{p-1}\frac{1}{a_\alpha^{p+1}}\ne 0.
\ea
Then,  the expectation values of the characteristic polynomials lead to (\ref{generalized}) in the double scaling limit.

We  first consider the intersection numbers with one marked point.
They are related to  the coefficients
of $\tr \frac{1}{\Lambda^m}$, in the zero-replica limit  $k\rightarrow 0$ as we have seen
in the previous section.
In this limit the matrix $\Lambda$
can be taken as multiple of the identity $\Lambda = \lambda\cdot {\bf 1}$.
We introduce a coupling constant $g$ as
\be
B \rightarrow \frac{B}{g}
\ee
\be
Z_p= \int dB e^{\frac{1}{(p+1)g^{p+1}} \tr B^{p+1} - 
\frac{\Lambda}{g}\tr B}
\ee
We set $g_s = g^{p+1}$ and tune $\Lambda$ so that 
$g_s = \frac{g}{\Lambda}$.
Then we obtain, after  the shift $B \to 1+ B$,
\be\label{log}
Z_p= \int dB e^{\frac{1}{(p+1)g_s} \tr (1+B)^{p+1} 
-\frac{1}{g_s}\tr (1+B)}
\ee
Expanding for small $B$, we have
\be
Z_p = \int dB e^{\frac{p}{2 g_s} \tr B^2 
+\frac{p(p-1)}{3! g_s}\tr B^3 +
\frac{p(p-1)(p-2)}{4! g_s} \tr B^4 \cdots}
\ee
We now expand in powers of $g_s$ 
after the replacement $B\to i\sqrt{\frac{g_s}{p}} B$.

Using the replica formula  (\ref{replica}) for one marked point, we obtain
$\lim_{k\to 0} \frac{1}{k}<\tr B^4> = 1, \lim_{k\to 0} \frac{1}{k}
<(\tr B^3)^2> = 3,...$. 
Using these values for the products of vertices, we have
\be\label{logZ}
\lim_{k\to 0} {\rm log} Z_p = - \frac{p-1}{24} (\frac{g_s}{p}) + 
\frac{(p-3)(p-1)(p+3)(1+2 p)(3+ 2 p)}{1920} \frac{g_s^3}{p^3} + O(g_s^5)
\ee

The coefficients of the above expansion are intersection numbers multiplied by
$t_{m,j}$ in (\ref{coefficientt}),
the first term for genus one,  the second  for
genus two, etc.. 
From (\ref{coefficientt}), we have 
$t_{1,0}= -\frac{1}{p} {\rm tr} \frac{1}{\Lambda^{p+1}}$.
Therefore, the intersection number of one marked point for genus one becomes
\be
<\tau_{1,0}>_{g=1} = \frac{p-1}{24}.
\ee

For genus two, we have $t_{3,2}= (-p)^{-2} 3 (3+p)(3+2 p){\rm tr}\frac{1}{\Lambda^{3p+3}}$, and 
\be\label{log32}
<\tau_{3,2}>_{g=2} = \frac{(p-1)(p-3)(1+ 2 p)}{p\cdot 5!\cdot 4^2\cdot 3}.
\ee
For $g=2$ and $p=2$, we have
$t_{4,0} = -\frac{3\cdot 5\cdot 7}{2^3}{\rm tr}\frac{1}{\Lambda^9}$ and
this gives $<\tau_{4,0}>_{g=2} = \frac{1}{1152}$. 
The expansion of (\ref{logZ}) can be obtained for any higher order of 
genus
$g$ by the use of replica formula of (\ref{replica}) 
although the evaluation becomes tedious.

We now turn to the dual model, formulated with $N\times N$ random matrices $M$ ;  the Fourier
transform $U(s)$ of the one point correlation function is given in  
(\ref{U(sn)}).

We are still in  the case in which the external source 
$a_\alpha$ takes values $p-1$ different
values of $a_1,a_2,...,a_{p-1}$ with $\frac{N}{p-1}$ degeneracy.

We have from (\ref{U(sn)}),
\be
U(s) = \frac{e^{\frac{s^2}{2}}}{Ns}\oint \frac{du}{2 i\pi}e^{u s}
e^{\sum_{\alpha=1}^{p-1}\frac{N}{p-1} \log (a_\alpha - u - s)-
\sum_{\alpha=1}^{p-1}\frac{N}{p-1} \log(a_\alpha - u)}
\ee
Expanding $u(s)$ for small $u$ and $s$, we have
\be
U(s) = \frac{e^{\frac{s^2}{2}}}{Ns}\oint \frac{du}{2 i\pi} 
e^{- s \sum \frac{N}{(p-1)a_\alpha}
 +(\frac{s^2}{2}+ u s)(1-
\frac{N}{p-1}\sum \frac{1}{a_\alpha^2}) 
- \sum_{n=3}^\infty
\frac{N}{n (p-1)a_\alpha^n}((u+s)^n- u^n)}
\ee

With the conditions (\ref{acond}),
we obtain,  after the shift $u\to u-\frac{s}{2}$,
\be\label{duallog}
U(s) = \frac{1}{Ns}\int \frac{du}{2i\pi} e^{-\frac{c}{p+1}
[
(u+\frac{1}{2}s)^{p+1} - (u - \frac{1}{2}s)^{p+1}]}
\ee
where $c= \frac{N}{p-1}\sum \frac{1}{a_\alpha^{p+1}}$.

Expanding the exponent, we obtain
\ba\label{Bexpansion}
&&U(s) =\frac{1}{Ns}\int \frac{du}{2 i \pi}{\rm exp}
[-cs u^p]\nonumber\\
&&\times {\rm exp}[-c(  \frac{p(p-1)}{3! 4} s^3 u^{p-2}
+ \frac{p(p-1)(p-2)(p-3)}{5! 4^2}s^5 u^{p-4}+\cdots)].
\ea

This integrals yield Gamma functions 
after the replacement 
$u= (\frac{t}{c s})^{1/p}$,
\ba\label{Mexpansion}
U(s) &=& \frac{1}{Nsp\pi}\cdot
\frac{1}{(c s)^{1/p}}\int_0^\infty dt
t^{\frac{1}{p}-1}e^{-t }\nonumber\\
&&\times e^{- \frac{p(p-1)}{3! 4} s^{2+\frac{2}{p}}
c^{\frac{2}{p}} t^{1 - \frac{2}{p}}
- 
\frac{p(p-1)(p-2)(p-3)}{5! 4^2} s^{4+\frac{4}{p}} c^{\frac{4}{p}}
t^{1 - \frac{4}{p}}+\cdots}\nonumber\\
%%%%%%%%%%%%%%%%%%%%%%%%%%%%%%%%%%%%%%%%%%%
&&= \frac{1}{Ns\pi}\cdot
\frac{1}{(c s)^{1/p}}
[ \Gamma(1+\frac{1}{p})- \frac{p-1}{24}  y \Gamma(
1 - \frac{1}{p}) \nonumber\\
&&+
\frac{(p-1)(p-3)(1+ 2 p)}{5! \cdot4^2\cdot3} 
y^2 \Gamma(1- \frac{3}{p})\nonumber\\
&&- \frac{(p-5)(p-1)(1+ 2 p)(8 p^2 - 13 p - 13)}{7! 4^3 3^2}
y^3 \Gamma(1 - \frac{5}{p}) \nonumber\\
&&+ (p-7)(p-1)(1+ 2 p) (72 p^4 - 298 p^3 - 17 p^2 + 562 p + 281)
\nonumber\\
&&\times \frac{1}{9! 4^4 15} y^4 \Gamma(1 - \frac{7}{p}) \cdots]
\ea
with
$y = c^{\frac{2}{p}}s^{2 + \frac{2}{p}}$.

Comparing this expansion with (\ref{logZ}), we
find  a common $p$-dendence, but (\ref{logZ}) has  additional
factors. These additional factors should be included
in the normalization.
The intersection numbers $<\tau_{n,j}>_g$ are the coefficients
of (\ref{Mexpansion}), and then the two results of (\ref{logZ}) 
and (\ref{Mexpansion}) coincide. The expansion of $U(s)$ is easily
obtained up to arbitrary order of genus $g$ since $U(s)$ is simply
given by (\ref{duallog}). Thus the dual model is simpler than the 
the partition function of the matrix $B$.

The expansion of $U(s)$ in (\ref{Mexpansion}) is genus expansion.
We write this expansion as
\be
U(s) = \sum_{g} <\tau_{n,j}>_{g} \frac{1}{N\pi} \Gamma(1 - \frac{1}{p}-\frac{j}{p})
c^{\frac{2g-1}{p}}p^{g-1}s^{(2g-1)(1+ \frac{1}{p})},
\ee
where $n$ and $j$ are given by 
\be\label{condone}
(p+1)(2 g -1) = p n + j + 1.
\ee
From (\ref{Mexpansion}), the intersection numbers $<\tau_{n,j}>$ are determined
explicitly for arbitrary $p$.

For $g=1$ case, by the condition of (\ref{condone}), we obtain $n=1$ and $j=0$,
and
\be\label{gg1}
<\tau_{1,0}>_{g=1} = \frac{p-1}{24}.
\ee

For $g=2$, we have $n = 3+ \frac{2-j}{p}$.  The intersection numbers
for arbitrary $p$  become
\be\label{gg2}
<\tau_{n,j}>_{g=2} = \frac{(p-1)(p-3)(1+ 2 p)}{p\cdot 5! \cdot 4^2\cdot 3}
\frac{\Gamma(1-\frac{3}{p})}{\Gamma(1- \frac{1+j}{p})}.
\ee
For instance, we obtain $<\tau_{3,2}>_{g=2} = \frac{(p-1)(p-3)(1+2p)}{p5!\cdot 4^2\cdot 3}$,
 which agrees
with the result of (\ref{log32}) evaluated from (\ref{logZ}). For $p=2$, we have from this formula,
$<\tau_{4,0}>_{g=2} = \frac{1}{(24)^2 2!}$, since the ratio of the gamma functions becomes
$-2$.

For $g=3$, we have  the intersection numbers for arbitrary $p$ with $n=5+ \frac{4-j}{p}$,
\be\label{gg3}
<\tau_{n,j}>_{g=3} = \frac{(p-5)(p-1)(1+ 2 p) ( 8p^2 - 13 p - 13)}{p^2\cdot
7!\cdot 4^3\cdot 3^2}\frac{\Gamma(1 - \frac{5}{p})}{\Gamma(1 - \frac{1+j}{p})}.
\ee

For $g=4$, we have  the intersection numbers with thecondition $n= 7+ \frac{6-j}{p},$
\ba\label{gg4}
<\tau_{n,j}>_{g=4} &=& \frac{(p-7)(p-1)(1+2p)(72 p^4-298 p^3 -17 p^2 +562 p + 281)}{p^3\cdot 
9!\cdot
4^4\cdot 15}\nonumber\\
&\times& \frac{\Gamma(1- \frac{7}{p})}{\Gamma(1 - \frac{1+j}{p})}.
\ea

Thus the expansion of $U(s)$ gives the intersection numbers of one marked point
 for arbitrary $p$ in the case of  fixed $g$. We have given the explicit expressions
up to order $g=4$ only.
Also the integral representation of $U(s)$ in (\ref{duallog}) 
gives the intersection numbers for arbitrary
$g$ for fixed $p$. As shown before in \cite{BH3}, the intersection numbers for $p=2$ becomes
\be
<\tau_{n,0}>_g = \frac{1}{(24)^g g!}, \hskip 5mm ( n= 3g -2 ).
\ee
For $p=3$, 
\be
<\tau_{n,j}>_g = \frac{1}{(12)^g g!}\frac{\Gamma(\frac{1+g}{3})}{
\Gamma(\frac{2-j}{3})},
\ee
 with
$3n = 8g - 5-j$. 

We now consider the interesting limit $p\to -1$ . We consider the spin index $j$ as $j=0$
in this limit. From the condition of (\ref{condone}), we obtain $n=1$. The intersection number
of $p=-1$ case is then $<\tau_{1,0}>_g$, which we write simply as $<\tau>_g$ in the following.
From the previous evaluations in (\ref{gg1})-(\ref{gg4}), 
we obtain in the limit $p\to -1$,
\ba\label{p-1}
&&<\tau>_{g=1} = \frac{p-1}{24} \to -\frac{1}{12}\nonumber\\
&&<\tau>_{g=2} = \frac{(p-1)(p-3)(1+ 2 p)}{5! 4^2} \frac{\Gamma(
1 -\frac{3}{p})}{\Gamma(1 - \frac{1}{p})}\to - \frac{1}{120}\nonumber\\
&&<\tau>_{g=3} \to - \frac{1}{252},\hskip 6mm <\tau>_{g=4}\to - \frac{1}{240}.
\ea
These numbers are the Euler characteristics $\chi(M_{g,1})$ \cite{Penner,HZ}. 
\be\label{chizeta}
\chi(M_{g,1}) = \zeta(1 - 2 g) = -\frac{B_{2g}}{2g}
\ee
where $\zeta$ is the Riemann zeta-function and 
$B_{2g}$ is the Bernoulli
number; $B_{2}= \frac{1}{6}, 
B_{4}=-\frac{1}{30}, B_6=\frac{1}{42}, B_8= \frac{1}{30} \cdots$.

The logarithmic term of the Penner model follows indeed in the limit
$p\to -1$ of (\ref{duallog}). Then computing $U(s)$ for $p\to -1$, one obtains the  
Euler characteristics.

In the limit $p\to -1$, $c= \frac{N}{p-1}\sum \frac{1}{{a_\alpha}^{p+1}}$ is $N$,
and
from (\ref{duallog}), $U(s)$ is given by
\ba\label{Eu}
U(s) &=&\frac{1}{Ns} \int \frac{du}{2i \pi} 
e^{-N \log \frac{u+\frac{1}{2}s}{u -\frac{1}{2}s}}\nonumber\\
&=&\frac{1}{N}\int \frac{du}{2 i \pi}
(\frac{u-\frac{1}{2}}{u+\frac{1}{2}})^N
\ea

Setting 
\be\label{uy}
\frac{u-1}{u+1}= e^{-y},
\hskip 3mm (u= \frac{1+e^{-y}}{1- e^{-y}})
\ee
one has 
\be
U(s) = -  \frac{1}{N}\int \frac{dy}{2\pi} \frac{e^{-y}}{(1- e^{-y})^2}
e^{- N y} =  \int_0^\infty \frac{dy}{2\pi} \frac{e^{-Ny}}{1- e^{-y}} 
\ee
Noting that
\be
\frac{1}{1 - e^{-t}} = \sum_{n=0}^\infty
B_n \frac{t^{n-1}}{n!}
\ee
we obtain $U(s)$ as 
\ba
&&U(s) = \frac{1}{N}\int_0^\infty dt \frac{1}{1 - e^{-\frac{t}{N}}} e^{- t}
= \sum_{n=0}^\infty \frac{B_n}{n}(\frac{1}{N})^{n}\nonumber\\
&&= 1 - \frac{1}{2N} + \frac{1}{12 N^2} - \frac{1}{120}\frac{1}{N^4}
+ \frac{1}{252}\frac{1}{N^6} + \cdots]
\ea
Then we obtain the genus $g$,  orbifold Euler characteristics 
$\chi(M_{g,1}) = \zeta(1 - 2g)= -\frac{1}{2g} B_{2g}$ from the term of order
$1/N^{2g}$.
Thus the analytic continuation for negative $p$ holds
for the dual model. 

\section{The n-point correlation functions}

We consider the  two-point correlation function $U(s_1,s_2)$ defined in (\ref{U1m}).
Noting that the two terms of the determinant in (\ref{U(sn)}) become, after the shift
$u_i\to u_i-\frac{s_i}{2}$, $s_i\to \frac{s_i}{N}$,
\ba
&&\frac{1}{u_1-u_2 + \frac{1}{2N}(s_1+s_2)} \frac{1}{u_1-u_2- \frac{1}{2N}(s_1+s_2)}\nonumber\\
&&= (\frac{1}{u_1-u_2 - \frac{1}{2N}(s_1+s_2)} - \frac{1}{u_1-u_2+ \frac{1}{2N}(s_1+s_2)})
\frac{N}{s_1+s_2}
\ea
we write it as
\ba
&&\frac{1}{u_1-u_2 + \frac{1}{2N}(s_1+s_2)} \frac{1}{u_1-u_2- \frac{1}{2N}(s_1+s_2)}\nonumber\\
&&= \frac{N}{s_1+s_2} \int_0^\infty dx e^{-x(u_1-u_2)}{\rm sh} (\frac{x}{2N}(s_1+s_2))
\ea
We have at the same p-th critical point defined for the  one point function,
\ba
&&U(s_1,s_2) = \frac{2N}{s_1+s_2}\frac{1}{(2\pi i)^2}\int_0^\infty dx
\int du_1 du_2 {\rm sh}(\frac{1}{2N} x (s_1+s_2))\nonumber\\
&&\times{\rm exp}[ - \frac{N}{p^2-1}\sum \frac{1}{a_\alpha^{p+1}} 
(\sum_i (u_i + \frac{1}{2N}s_i)^{p+1}
-\sum_i (u_i-\frac{1}{2N}s_i)^{p+1}) 
\ea
We use the notation, $ c= \sum_\alpha \frac{1}{a_\alpha^{p+1}}$. After the change of variables
$u_i \to i v_i$ (i=1,2), the rescalings $v_i\to (\frac{p-1}{pc s_i})^{\frac{1}{p}} v_i$, and 
$x\to (\frac{pc s_1}{p-1})^{\frac{1}{p}}x$,
 we obtain 
\ba\label{utwo}
&&U(s_1,s_2) = \frac{2N'}{s_1+s_2} (\frac{1}{s_2})^{\frac{1}{p}}
\int_0^\infty dx \int_{-\infty}^\infty
\frac{dv_1 dv_2}{(2\pi)^2} {\rm sh} (\frac{x}{2N'} s_1^{\frac{1}{p}} (s_1+s_2))\nonumber\\
&&e^{-i x v_1 + i x v_2 (\frac{s_1}{s_2})^{\frac{1}{p}}} \prod_{i=1}^2 G(v_i)
\ea
where we used $N'= N (\frac{p-1}{pc})^{\frac{1}{p}}$, and $[\frac{p}{2}]=\frac{p}{2}$ for
even p and $[\frac{p}{2}]=\frac{p-1}{2}$ for odd p.
The factor $G(v_i)$ is given by
\be\label{G(v)}
G(v_i) ={\rm exp}
[ - \frac{i^p}{p}v_i^p - i^p \sum_{m=1}^{[\frac{p}{2}]}\frac{(-1)^m (p-1)!}{(2m+1)! 2^{2m}(p-2m)! N'^{2m}}
s_i^{(2+\frac{2}{p})m} v_i^{p-2m}].
\ee
The genus $g$ of the terms in the expansion  (\ref{G(v)}) 
is given  by the  exponent of  $\frac{1}{N'^{2g}}$.
We are interested in the terms of type $s_1^{n_1+\frac{m_1}{p}} s_2^{n_2+\frac{m_2}{p}}$ in (\ref{utwo}).
The correspondence with the variable $t_{n,m}\sim {\rm tr} \frac{1}{\Lambda^{pn+m+1}}$ is
\be
s^{n+\frac{m+1}{p}} \sim t_{n,m}
\ee
Thus we obtain the intersection numbers $<\tau_{n_1,m_1}\tau_{n_2,m_2}>_g$
from the coefficients of the terms $s_1^{n_1+\frac{m_1+1}{p}}s_2^{n_2+\frac{m_2+1}{p}}$.
For instance, we have from $s_1^{\frac{1}{p}}s_2^{2+\frac{1}{p}}$ in (\ref{utwo}),
\be
<\tau_{0,0}\tau_{2,0}>_{g=1} = \frac{p-1}{24}
\ee
This value coincides with $<\tau_{1,0}>_{g=1}= \frac{p-1}{24}$, and
we have
\be
<\tau_{0,0}\tau_{2,0}>_{g=1} = <\tau_{1,0}>_{g=1}
\ee
which is consistent with the string equation. Indeed the generating function
$F$ for the intersection numbers satisfies the  string equation \cite{Witten1},
\be\label{stringequation}
\frac{\partial F}{\partial t_{0,0}} = \frac{1}{2}\sum_{m,m'=0}^{p-2}
\eta^{m m'} t_{0,m} t_{0,m'} + \sum_{n=1}^\infty \sum_{m=0}^{p-2} t_{n+1,m}\frac{\partial F}{
\partial t_{n,m}}
\ee
where the metric $\eta^{m m'}= \delta_{m+m',p-2}$. 

If we substitute $F= a t_{1,0}$ ($a$ is some constant) in the second
term, then we have $\frac{\partial F}{\partial t_{0,0}}= t_{2,0}\frac{\partial F}{\partial
t_{1,0}}= a t_{2,0}$, which yields $a t_{0,0}t_{2,0}$ after integration.
Thus we have $<\tau_{0,0}\tau_{2,0}> = <\tau_{1,0}>$ from the string equation (\ref{stringequation}).

Note that in the two point correlation (two marked points), there is no genus zero
contribution to the intersection numbers, an easy consequence of (\ref{utwo}).

It is useful to define the higher Airy functions $\phi_p(x)$ by 
\be
\phi_p(x) = \int \frac{dv}{2\pi} e^{-\frac{i^p}{p}v^p + i x v}
\ee
When $p=3$, it reduces to the usual Airy function $A_i(x)$,
\be
\phi_3(x)=A_i(x)= \frac{1}{2\pi} \int_{-\infty}^\infty dv e^{\frac{i}{3}v^3 + i x v}
\ee
which satisfies the differential equation,
\be
\phi_3(x)^{\prime\prime} = x \phi_3(x)
\ee
For general $p$, we have
\be
\frac{d^{p-1} \phi_p(x)}{dx^{p-1}} = x \phi_p(x)
\ee
Expanding in powers of $\frac{1}{N'}$, after integration by parts for
the function $\phi_p(x)$, we obtain all intersection numbers from (\ref{utwo}) for two 
marked points.
In the case $p=2$, the function $\phi_p(x)$ is  Gaussian , and
we  are led to
\ba
&&U(s_1,s_2) = \frac{2N}{(s_1+s_2) \sqrt{s_2}}e^{\frac{1}{24 N^2}(s_1^3+s_2^3)}\int_0^\infty
dx \hskip 1mm{\rm sh}(x\frac{\sqrt{s_1}}{2N}(s_1+s_2))
e^{-\frac{1}{2}\frac{s_1+s_2}{s_2}x^2}\nonumber\\
&& = \frac{N}{s_1+s_2} e^{\frac{1}{24 N^2}(s_1+s_2)^3} \sum_{m=0}^\infty
\frac{(-1)^m}{m! (2m+1)}(\frac{s_1s_2(s_1+s_2)}{8N^2})^m \sqrt{s_1 s_2}
\ea
which has been obtained in \cite{BH1,Okounkov}. The intersection numbers are given by
\be
U(s_1,s_2) = \sum_{n_1,n_2} <\tau_{n_1,0}\tau_{n_2,0}>_g \frac{s_1^{n_1}s_2^{n_2}}{N^{2g}}
\ee
where the genus $g$ is specified by
\be
n_1+n_2 = 3g-2.
\ee

For the  $p=3$ case, we have
\ba
U(s_1,s_2) = \frac{2 N'}{s_1+s_2}(\frac{1}{\sqrt{s_2}})^{\frac{1}{3}}\int_0^\infty dx
{\rm sh}(\frac{x}{2 N'}s_1^{\frac{1}{3}}(s_1+s_2)) A_i(x)A_i(-x(\frac{s_1}{s_2})^{\frac{1}{3}})
\ea
which gives the intersection numbers of two-marked points. For instance, we obtain
\be
<\tau_{0,0}\tau_{2,0}>_{g=1}= \frac{1}{12}, \hskip 5mm <\tau_{1,0}^2>_{g=1}= 
\frac{1}{12}
\ee

In general, the integral formula of (\ref{U(sn)}) gives the  n-point correlation function $U(s_1,...,s_n)$.
In Appendix A,  we compute the intersection numbers for three marked points  from the three point function$U(s_1,s_2,s_3)$. In Appendix B, the intersection numbers with four marked points are computed from the
four point correlation function.

The intersection numbers for the primary field $U_j$ of (\ref{Uj}),  in the genus zero case,  are particularly important,
since they have algebraic structures related to a  super-conformal field theory \cite{Witten1, Dijkgraaf1}. They are expressed as
$<\prod_{m=1}^n \tau_{0,q_m}>$.

In Appendix A, we compute the intersection numbers  (\ref{3point}) as
\be
< \tau_{0,q_1}\tau_{0,q_2}\tau_{0,q_3}>_{g=0} = \delta_{q_1+q_2+q_3,p-2}
\ee
From this result, the free energy $F$ follows
\be
F = \sum <\tau_{0,q_1}\tau_{0,q_2}\tau_{0,q_3}>_{g=0} t_{0,q_1}t_{0,q_2}t_{0,q_3} +O(t^4)
\ee
To make the algebraic structure more explicit, we define the structure constants $C_{ijk}$  by
\be\label{structure}
C_{ijk} = \frac{\partial^3 F}{\partial t_i \partial t_j \partial t_k}
\ee
where 
\be
t_i = t_{0,i-1}, \hskip 5mm (i=1,...,p-1)
\ee
For instance for p=5, we have
\be
F = \frac{1}{2} t_{0,0}^2 t_{0,3} + t_{0,0}t_{0,1}t_{0,2} + \frac{1}{3!}t_{0,1}^3 + O(t^4)
\ee
and the structure constants are
$C_{114}=C_{123}=C_{222}=1$.
From these structure constants,  following \cite{Witten1} one can construct the super potential $W$ (see appendix C).
The free energy $F$, which 
generates the intersection numbers of the moduli space of p-spin curves, has been conjectured by Witten to be solution of the
Gelfand-Dikii hierarchy \cite{Witten1}. We present in Appendix C, this Gelfand-Dikii hierarchic
equations as well as  the
construction of the super potential for the primary fields.

Thus, we find that the intersection numbers, 
derived  from the integral representation
of $U(s_1,...,s_n)$, satisfy indeed the Gelfand-Dikii equations.

Returning to  Witten's conjecture, we note that the definition of the intersection numbers by
the vector bundle integration over the compactified moduli space $\bar M_{g,n}$ in (\ref{Uj}), is 
similar in structure to the integral representation for $U(s_1,...,s_n)$ of 
(\ref{U(sn)}), although
$U(s_1,...,s_n)$ involves a summation over all genuses  $g$, for fixed n-marked points.

\section{Time-dependent Gaussian matrix model}

Let us  briefly recall how one shows that the time-dependent one matrix model is
equivalent to a time-independent two-matrix model when the distributions are Gaussian. 
( In an older work we had considered the time-dependent Gaussian matrix problem, 
and computed time-dependent correlation functions \cite{BH6}).

The time-dependent Gaussian matrix model is the partition function 
(i.e. the Euclidean path integral) for the matrix quantum mechanics, with action
\be
S = \int dt \frac{1}{2}{\tr} (\dot{M}^2  + M^2)
\ee
where $M$ is  an $N\times N$ Hermitian matrix ; (the dot stands for  time derivative). 
This model, at criticality, is known to describe gravity coupled to matter 
of central charge $c=1$ \cite{BKZ}. 

The time-dependent correlation function is defined by
\be
\rho(\lambda,\mu;t) = < \frac{1}{N}{\rm tr} \delta(\lambda-M(t_1))
\frac{1}{N}{\tr} \delta(\mu - M(t_2))>
\ee
and, from time-translation invariance, is function of $t= |t_2-t_1|$. 
The Fourier transform of this quantity is
\be
U(\alpha,\beta) = \frac{1}{N^2}<{\tr} e^{i\alpha M(t_1)}{\tr}e^{i\beta M(t_2)}>
\ee
This correlation function may easily be reduced to the correlation function of 
the time-independent two matrix model in the
Gaussian ensemble \cite{BH6}.
Indeed this harmonic oscillator quantum mechanics leads to 
\ba
U(\alpha,\beta) &=& \frac{1}{N^2}(\frac{e^t}{{\rm sinh} t} )^{N^2/2}\int 
dA dB \tr e^{i \alpha A} \tr
e^{i\beta B}\nonumber\\
&\times& e^{-\frac{1}{2}{\rm sh} t \tr [(A^2+B^2)e^t - 2 A B]}
\ea
Rescaling of A,B,$\alpha$ and $\beta$ by a factor $\sqrt{e^{-t}{\rm sinh}t}$, we obtain
a  time-independent two-matrix model 
\be
U(\alpha,\beta) =\frac{1}{Z} \int dA dB {\tr} e^{i\alpha A} \tr e^{i\beta B} e^{-\frac{1}{2}{\tr}(A^2+B^2-2c A B)},
\ee
with a coupling constant $$c= e^{-t} .$$

For convenience, we denote $A$ and $B$ by $M_1$ and $M_2$ in the following .

\section{Duality formula for the two-matrix model}

We consider the correlation function of the characteristic polynomials 
in the two-matrix model,
\be\label{J}
J = <\prod_{\alpha=1}^{k_1}{\rm det}(\lambda_\alpha - M_1)
\prod_{\beta=1}^{k_2}{\rm det}(\mu_\beta - M_2)>
\ee
where the average  is perfomed over a two-matrix Gaussian distribution with 
an external source $A$ acting on one of the two $N\times N$ matrices,
\be
P(M_1,M_2)= \frac{1}{Z} e^{-\frac{1}{2}{\tr} M_1^2 -\frac{1}{2}{\tr} M_2^2 - 
c {\tr} M_1 M_2 - {\tr} M_1 A}
\ee
The external source $A$ will be used here again  mean 
to tune a   $p$-spin structure in the moduli space.
Note that when $t\to \infty$  
the parameter $c=e^{-t}$ vanishes and  the two matrices $M_1$ and $M_2$ decouple.

From (\ref{J}) we shall determine a new dual model of Kontsevich type in the large N limit.
The duality formula for $J$ is obtained  by the use of Grassmann variables  
as in the one matrix case \cite{BH2,BH4}, but 
for the two matrix case a new structure appears. 

Let us introduce the Grassmann variables $\psi_i^{\alpha}$ and $\chi_i^{\beta}$,
where $\alpha=1,...,k_1$, and $\beta=1,...,k_2$. Then

\be
J = < \int d\bar\psi d\psi d\bar \chi d \chi e^{N[\bar\psi_\alpha(\lambda_\alpha - M_1)\psi_\alpha
+ \bar\chi_\beta(\mu_\beta - M_2)\chi_\beta]} >
\ee
Since the probability $P$ is  Gaussian, one can integrate out the matrices $M_1$ and $M_2$. This generates four-fermion terms that may be disentangled with the help of three auxiliary matrices  : $B_1$  a $k_1\times k_1$ Hermitian matrix, $B_2$  a $k_2\times k_2$ Hermitian matrix  and $D$ a complex $k_1\times k_2$ rectangular matrix. The identities

\be
e^{-\frac{N}{2(1-c^2)}\bar\psi\psi\bar\psi \psi} = 
\int dB_1 e^{-\frac{N}{2}{\tr}B_1^2 + \frac{iN}{\sqrt{1 - c^2}}{\tr}
B_1\bar \psi \psi}
\ee
\be
e^{-\frac{N}{2(1-c^2)}\bar\chi\chi\bar\chi \chi} = \int dB_2 e^{-\frac{N}{2}{\tr}B_2^2 + \frac{iN}{\sqrt{1 - c^2}}{\tr}
B_2\bar \chi \chi}
\ee
\be
e^{\frac{Nc}{1-c^2}\bar\psi\chi\bar\chi\psi} =
\int dD dD^{\dagger}
e^{-N {\tr}D^{\dagger}D + N\sqrt{\frac{c}{1-c^2}}{\tr}(D\bar\psi \chi + D^{\dagger}\bar\chi \psi)}
\ee
allow to represent $J$ as 
\ba
&&J= \int dB_1 dB_2 dD^{\dagger} dD e^{-\frac{N}{2}{\tr}(B_1^2+B_2^2+ 2 D^{\dagger} D)}\nonumber\\
&&\times \prod_{i=1}^N {\rm det} \left( \matrix{
(\lambda_\alpha  - \frac{a_i}{1-c^2})\delta_{\alpha,\alpha'} + \frac{i}{\sqrt{1-c^2}} B_1 &
\sqrt{\frac{c}{1-c^2}} D\cr
\sqrt{\frac{c}{1-c^2}}D^{\dagger} & (\mu_\beta + \frac{c}{1-c^2} a_i)\delta_{\beta,\beta'} + \frac{i}{\sqrt{1-c^2}} B_2\cr}
\right)\nonumber\\
\ea
After the shift $B_1\rightarrow B_1+ i\sqrt{1-c^2} \lambda_{\alpha,\alpha'}\delta_{\alpha,\alpha'}$ and
$B_2\rightarrow B_2+ i\sqrt{1-c^2} \mu_{\beta,\beta'}\delta_{\beta,\beta'}$,
we obtain the dual expression for $J$ : 
\ba\label{X}
J &=& C \int dB_1 dB_2 dD^\dagger dD
e^{-\frac{N}{2}{\tr}(B_1^2+B_2^2+ 2 D^\dagger D) - i N \sqrt{1-c^2}{\tr}B_1\Lambda_1
- i N \sqrt{1 - c^2}{\tr}B_2 \Lambda_2}\nonumber\\
&\times& e^{-\sum_{i=1}^N {\tr} {\rm log}(1 - X_i)},
\ea
where the matrices $X_i$ are  defined by
\be\label{Xi}
X_i = \left( \matrix{ 
i\sqrt{1-c^2}\frac{B_1}{a_i} & 
\sqrt{c(1-c^2)} \frac{D}{a_i}\cr
- \frac{\sqrt{c(1-c^2)}}{c} \frac{D^{\dagger}}{a_i} 
&- \frac{i\sqrt{1-c^2}}{c} 
\frac{B_2}{a_i}
\cr } \right).
\ee
We now expand ${\rm log}(1- X_i)$ in powers of $X_i$.
Let us first consider 
the case of a source matrix A mutiple of the identity :   $a_i=a, \ i=1 \cdots N$. 
Imposing the constraint \be
a= (1-c^2)
\ee
the quadratic term  ${\tr}B_1^2$ cancels with the one coming from  the expansion 
of ${\rm log}(1- X)$. This critical constraint corresponds to the edge of the
the spectrum for the matrix $M_1$. 
Note that the $B_2^2$ term is not cancelled at this critical point because of the coupling $c$. 

Given the  factor $N$
in the exponent, the edge scaling limit under consideration corresponds  to \be\label{order}
B_1\sim O(N^{-\frac{1}{3}}), B_2\sim O(N^{-\frac{1}{2}}), D\sim O(N^{-\frac{1}{3}})
\ee
in the large N limit.
In this limit most terms disappear ; for instance
\be
N {\tr} (D^\dagger D B_2)\sim N^{-\frac{1}{6}}
\ee
is negligible.
Then, in the large N limit (\ref{order}), we obtain the partition function $Z$, i.e. $J$ after dropping the negligible terms,
\be
Z= \int dB_1 dB_2 d D^\dagger d D e^{- i N {\tr} B_1 \Lambda_1 - i N {\tr} B_2 \Lambda_2
+ \frac{i}{3}N {\tr} B_1^3 - \frac{N}{2}(1- \frac{1}{c^2}) {\tr} B_2^2 + i N {\tr}
(D D^\dagger B_1)}
\ee
Since the matrix $B_2$ matrix is decoupled we can  integrate it out. Then, dropping the decoupled part, 
we find the  partition function   
\be
Z = \int dB_1 dD^\dagger dD e^{-i {\tr}B_1 \Lambda_1 + \frac{i}{3}{\tr} B_1^3 + i {\tr} D D^\dagger B_1}\ee
where we have absorbed the powers of  $N$ given by  the scaling (\ref{order}).
We may now  integrate out the matrices $D$ and $D^\dagger$ ; this yields a one matrix integral with a  logarithmic potential,
\be\label{log}
Z = \int dB_1 e^{\frac{i}{3}{\tr} B_1^3 - k_2 {\tr}{\rm log} B_1 - i {\tr} B_1 \Lambda_1}.
\ee
The appearance of a logarithmic term is a  characteristic of models with  
central charge equal to one. 

We now consider the free energy of this logarithmic 
Kontsevich model (p=2) (\ref{log}). 
Three different, but consistent, methods will be used. 
(For convenience   $k_2$ is denoted as $q$ in what follows.)

\vskip 5mm
{\bf i) HarishChandra-Itzykson-Zuber method}
\vskip 5mm
After use of the HarishChandra-Itzykson-Zuber formula, the partition function $Z$ is given by
\ba\label{60}
Z &=& \int dB e^{\frac{i}{3}{\tr} B^3 - q {\tr} {\rm log} B - i 
{\tr} B \Lambda^2}\nonumber\\
&=& \frac{1}{\Delta(l^2)}\int \prod_{i=1}^{k_1} 
dx_i \Delta(x) \prod_{i=1}^{k_1} e^{- i x_i l_i^2
+ \frac{i}{3}x_i^3 - q {\rm log} x_i}
\ea
where the $x_i$'s are the eigenvalues of $B$,  the $l_i$ the eigenvalues of $\Lambda$, and $\Delta(x)$ the Vandermonde determinant $\Delta(x)= \prod_{i<j} (x_i-x_j)$. It may be replaced in the 
integrand  by the Vandermonde of differential operators $\frac{\partial}{\partial l_i^2}$,
\be
Z= \prod \frac{1}{(l_i^2 -l_j^2)}(\frac{\partial}{\partial l_i^2}-\frac{\partial}{\partial l_j^2})
\prod \zeta(l_i)
\ee
where
\be
\zeta(l_i) = \int dx e^{\frac{i}{3} x^3 - i x l_i^2 - q {\rm log} x}.
\ee
Rescaling $x_i\rightarrow x_i/2^{1/3}$ and $l_i\rightarrow l_i/2^{1/3}$, and with the
change  $l\rightarrow i l$, we have
\be
\zeta(l) = \frac{e^{-\frac{1}{3}l^3}}{l^{q+\frac{1}{2}}}\int dx
e^{-\frac{1}{2}x^2 + \frac{i}{6 l^{3/2}}x^3 - q {\rm log}(1+ \frac{x}{i l^{3/2}})}
\ee
Expanding for large $l$, we obtain
\ba\label{texpansion}
{\rm log} Z &=& - [\frac{1}{6}t_1^3 + \frac{1}{24}t_3
+ q t_2 t_1+\frac{1}{2} q^2 t_3]
\nonumber\\
&&+ [ \frac{1}{6} t_1^3 t_3 + \frac{1}{48} t_3^2 + \frac{1}{8} t_1 t_5+ \frac{2}{3} q t_6 + q t_1^2 t_4\nonumber\\
&&+ q t_1 t_2 t_3 + \frac{1}{6}q t_2^3 + \frac{3}{2} q^2 t_1 t_5 + \frac{1}{4}q^2 t_3^2
+ q^2 t_2 t_4 \nonumber\\
&&+ \frac{2}{3} q^3 t_6] + O(\frac{1}{l^9})
\ea
where we have used the moduli parameters
\be
t_n = \sum_{i=1}^{k_1} \frac{1}{l_i^n}
\ee
When $q\rightarrow 0$, we recover the result of the one-matrix Kontsevich model.

For the relation to the genus $g$, we have to identify the powers of $\frac{1}{N^2}$. In the limit in which 
\be
k_1 \sim q \sim N, l_i \sim N^{\frac{1}{3}}
\ee
we find
\be
t_1 \sim O(N^{\frac{2}{3}}), t_n \sim O(N^{1-\frac{1}{3}n})
\ee
The genus expansion of the free energy 
\be
{\rm log}Z = \sum_{g=0}^\infty a_{g} N^{2- 2 g}
\ee
follows from this limit. For instance, $t_1^3$,$q t_1 t_2$, and  $q^2 t_3 $ are contributions to genus
zero, and $t_3$  to genus one.

\vskip 5mm
{\bf ii) replica method}
\vskip 5mm
We return to the integral (\ref{60}) .  After the  shift 
$B\to B+\Lambda$, which eliminates the terms linear in $B$, one can expand for large $\Lambda$. 
Then the  logarithmic potential, $\tr{\rm log}(B+ \Lambda)$ expanded in powers of  $\Lambda^{-1}$  yields  $\tr B^n$ vertices. The situation is similar to that of the generalized Kontsevich model where the
$\tr B^{p+1}$ terms led to the moduli space of  p-th curves, and spin structures appeared. 
The occurence  of
$t_2$,$t_4$ and $t_5$ indicates this fact.

We consider the moduli space for Riemann surfaces with  marked points. 
Although there is a logarithmic potential,
the model allows one to consider marked points, whose number is equal 
to the number of $t_n$.

We have developed the replica method $k_1\rightarrow 0$ in a previous article \cite{BH2}.
Any average of the products of vertices $\tr B^n$ are obtained in 
the replica limit $k_1\rightarrow 0$, where
$B$ is  a $k_1\times k_1$ Hermitian matrix ;   
for a Gaussian ensemble, the average is given by the  replica limit  formula
(\ref{replica}).

Expanding the logarithmic term, after the shift $B\rightarrow B + \Lambda$, and the
use of the formula (\ref{replica}), we
obtain the replica limit, which gives the required intersection numbers 
with one marked point. Indeed it leads easily  to
\be
{\rm log} Z = - ( \frac{1}{24} + \frac{1}{2} q^2) t_3 + (\frac{2}{3}q + \frac{2}{3}q^3) t_6 + O(\frac{1}{\Lambda^9})
\ee
This result agrees completely with the expression  (\ref{texpansion}) for one marked point.

\vskip 5mm
{\bf iii) differential equation of Virasoro type}
\vskip 5mm
Since the free energy ${\rm log}Z$ is expressed in 
terms of the moduli parameters $t_n$,  as was the case in the original Kontsevich model
(q=0 case),
it is  natural to investigate here again the KdV-like differential equations or string equations. Although there is a logarithmic potential
one may use  a Schwinger-Dyson equation \cite{Gross,Mironov}.

We first  consider the simple case, $k_1=1$, a one by one matrix, i.e. a c-number.
Denoting   $l_1= x$, one finds  
\be
Z = e^{-\frac{1}{3} x^3}\frac{1}{x^{q+\frac{1}{2}}}g(x) .
\ee

The Schwinger-Dyson (Virasoro) equation  follows from the identity
\be
\int dB \frac{\partial}{\partial B} e^{\frac{i}{3}{\tr B^3}-i {\tr B \Lambda^2} - q {\tr}{\rm log} B} = 0
\ee

The matrix $B$ is the replaced by $\frac{\partial}{\partial \Lambda^2}$. For the logarithmic potential, it means
$(\frac{\partial}{\partial \Lambda^2})^{-1}$. Therefore we need to apply a  differential operator in order to get rid of
this integral. We find easily that when $B$ is just a real number ($k_1=1$), the function $g(x)$ satisfies a third order differential equation,
\be
[ (\frac{\partial}{x\partial x})^3 - 2 + 2 q - x \frac{\partial}{\partial x}] \frac{e^{-\frac{1}{3}x^3}}{x^{q+ \frac{1}{2}}}
g(x) = 0
\ee
i.e.
\ba\label{diff}
&&(-(1+ 2 q)(5+ 2 q) (9 + 2 q) + (-10 -48 q - 24 q^2)x^3)g \nonumber\\
&&+ ((66  + 96 q + 24 q^2) x + (24 + 48 q ) x^4 + 16 x^7) g^{\prime}(x) \nonumber\\
&&+ (-36 x^2 -24q x^2-24 x^5)g^{\prime\prime}
+ 8 x^3 g^{\prime\prime\prime}(x) = 0.
\ea
This provides  the large $x$ expansion,
\ba
g(x) &=& -1 + \frac{1}{x^3}(\frac{5}{24} + q + \frac{q^2}{2})\nonumber\\
&-& \frac{1}{x^6} (  \frac{385}{1152} + \frac{73}{24} q + \frac{161}{48} q^2 + 
\frac{7}{6} q^3 + \frac{1}{8} q^4)
\nonumber\\
&+& \frac{1}{x^9}( \frac{85085}{82944} + \frac{6259}{384} q +
\frac{58057}{2304} q^2 + \frac{2075}{144} q^3
\nonumber\\
&& + \frac{725}{192} q^4 + \frac{11}{24} q^5 + \frac{1}{48} q^6) + O(\frac{1}{x^{12}}).
\ea
For $k_1=2$, a two by two matrix, we denote $x= l_1$ and $y=l_2$ ; then 
\be
[ (\frac{\partial}{x\partial x})^3 + \frac{\partial}{x\partial x}
(\frac{2}{x^2-y^2}(\frac{\partial}{x\partial x}-
\frac{\partial}{
y\partial y})) - 2 + 2 q - x \frac{\partial}{\partial x}] 
\frac{e^{-\frac{1}{3}(x^3+ y^3)}}{(x y)^{q + \frac{1}{2}} (x+y)}
g(x,y)=0
\ee
The solution, after symmetrization over $x$ and $y$, agrees with the expression (\ref{texpansion}).

When $x$ is small, the equation (\ref{diff}) leads to a different series expansion. The differential equation for 
$g$ in (\ref{diff}) has three different solutions,
\be
g(x) \sim x^{q + \frac{1}{2}}, g(x) \sim x^{q + \frac{5}{2}}, g(x)\sim x^{q+ \frac{9}{2}}
\ee
For small $x$, we have from the first solution, noting that $l= x$,
\be
Z = e^{-\frac{1}{3}l^3} ( 1+ \frac{1}{3} l^3 + (- \frac{1}{24}q + \frac{7}{72}) l^6 + O(l^9)
\ee
This solution for small $l$ is in a different phase from Kontsevich's phase ; it may be  related to one of the
two phases of the unitary matrix model \cite{Mironov,BG,GW}.

We now consider the case $p>2$. After the integration over the $D$-fields within the matrix $X$,
we also obtain the logarithmic term ${\rm tr log}B$, but corrections appear.
By tuning the external source with the conditions (\ref{acond}), we obtain
\be
Z = \int dX e^{-\frac{1}{p+1}{\rm tr} X^{p+1} + {\rm tr} X \Lambda^p}
\ee
where $X$ is given by
\be\label{X2}
X = \left( \matrix{ 
B & 
D\cr
 D^{\dagger}
& 0
\cr } \right).
\ee
where we have scaled out the factors $\sqrt{1- c^2}$ in $X_i$ of (\ref{X}), and put
$B=B_1$ and $B_2=0$.
We expand the potential,
\ba\label{Xp}
{\rm tr} X^{p+1} &=& {\rm tr}B^{p+1} + (p+1){\rm tr}D D^\dagger B^{p-1} + \frac{1}{2}
(p+1)(p-2) {\rm tr} (D D^\dagger)^2 B^{p-3}\nonumber\\
&&+ \frac{1}{6}(p+1)(p-3)(p-4){\rm tr} (D D^\dagger)^3 B^{p-5} + \cdots.
\ea
For $p=3$, we obtain,
\be\label{p=3D}
Z = \int dB dD^\dagger dD e^{  - 
[\frac{1}{4}{\tr} B^4 +
{\tr} D D^\dagger B^2
+ \frac{1}{2}{\tr}(D D^\dagger)^2] + {\rm tr} B \Lambda^3}.
\ee
The integration of the D-field can not be done explicitly for the
general values of $k_1$ and $k_2$ ($B$ is a $k_1\times k_1$ Hermitian matrix and $D$ is a 
$k_1\times k_2$ complex matrix). We make here a perturbation for the
large $B$ in lower orders. Expanding the term 
${\rm exp}(-\frac{1}{2}{\rm tr}( D D^\dagger )^2)$, we find
\be
Z = \int dB e^{+  {\tr} B \Lambda^3 - \frac{1}{4}
{\tr}B^4 - 2 k_2 {\tr}{\rm log}B - 
\frac{1}{2}k_2^2 ({\rm tr} \frac{1}{B^2})^2 -\frac{1}{2}k_2 
{\rm tr} \frac{1}{B^4} + O(\frac{1}{B^8}) }.
\ee
For $p=4$ case, 
the partition function $Z$ becomes similarly 
\ba
Z &=& \int dB dD dD^\dagger e^{
- \frac{1}{5}{\rm tr} B^5 + {\rm tr} D D^\dagger B^3 +
 {\rm tr} (D D^\dagger)^2 B + {\tr} B \Lambda^4}\nonumber\\
&=& \int dB e^{{\rm tr} B \Lambda^4-\frac{1}{5}{\rm tr} B^5 - 3 k_2 {\rm tr log}B - 
k_2 {\rm tr}\frac{1}{B^5} - k_2^2
{\rm tr} \frac{1}{B^2}{\rm tr}\frac{1}{B^3} + O(\frac{1}{B^{10}})}.
\ea

For general  $p$,  after  integration over the $D$-field in a perturbation, we obtain
\be\label{ZZZZ}
Z = \int dB e^{{\rm tr} B \Lambda^p -\frac{1}{p+1}{\rm tr}B^{p+1} - (p-1) k_2 {\rm tr log} B - 
\frac{p-2}{2}k_2 {\rm tr} \frac{1}{B^{p+1}} - \frac{p-2}{2} k_2^2 {\rm tr}\frac{1}{B^2} {\rm tr} \frac{1}{B^{p-1}}
+ \cdots}
\ee
In order to identify the intersection numbers, we expand in powers of $\frac{1}{\Lambda}$.
For this purpose, we  shift $B\to B+ \Lambda$. 
New terms in the exponent, which are corrections to the logarithmic term, have the form
of the product of two traces. In the large $\Lambda$ case, the fifth term is of the form
\be
{\rm tr}\frac{1}{(\Lambda + B)^2} {\rm tr} \frac{1}{(\Lambda + B)^{p-1}}\sim
-2 ({\rm tr} \frac{1}{\Lambda^3}B)\cdot {\rm tr}\frac{1}{\Lambda^{p-1}}+\cdots
\ee
The term ${\rm tr}\frac{1}{\Lambda^{p-1}}$ is $t_{p-1}$ ; such terms
appear in the $c=1$ string theory \cite{Mukhi1,Dijkgraaf2}.

We now evaluate the intersection numbers for a small number of marked points, and for lower
orders in  $\frac{1}{\Lambda}$. In this case, we can neglect the  above correction terms,
and we approximate the partition function by
\be
Z = \int dB e^{-\frac{1}{p+1}{\rm tr} B^{p+1} - (p-1)k_2 {\rm tr log} B + {\rm tr} B\Lambda^p}
\ee
We find, at order $\frac{1}{\Lambda^{p+1}}$, up to three marked points, 
\ba
{\rm log} Z &=& (\frac{p-1}{24})\frac{1}{p}\sum \frac{1}{{\lambda_i}^{p+1}}+
(\frac{p(p-1)}{12})\frac{1}{p}(\sum \frac{1}{\lambda_i})^2 (\sum \frac{1}{{\lambda_i}^{p-1}})\nonumber\\
&+& (\frac{p(p-1)}{2} k_2)\frac{1}{p}(\sum \frac{1}{\lambda_i^2})(\sum \frac{1}{{\lambda_i}^{p-1}})
+ (\frac{(p-1)^2}{2} k_2^2) \frac{1}{p}\sum \frac{1}{{\lambda_i}^{p+1}}\nonumber\\
&+& (higher \hskip 2mm order)
\ea
where the overall factor $\frac{1}{p}$ is a normalization constant absorbed in $\lambda$.
Thus we find the two-matrix model for $c=1$, obtained from the
characteristic polynomials, reduces to the one matrix model, and
the topological invariants becomes similar to the intersection numbers of 
$p$-spin curves.

When we put $p\to -1$, we find that the last term in (\ref{ZZZZ})
$
({\rm tr}\frac{1}{B^2})( {\rm tr} \frac{1}{B^{p-1}})$ behaves like
\be
({\rm tr}\frac{1}{B^2})( {\rm tr} \frac{1}{B^{p-1}}) = ({\rm tr}\frac{1}{B^2})({\rm tr} B^2)
\sim ({\rm tr}\frac{1}{\Lambda^2})({\rm tr} B^2)
\ee
where we make a shift $B \to \Lambda + B$.
In the case $p\to -1$,  when $B$ is order of $\Lambda$, all the terms of the potential
should be order of one, and indeed $t_m ({\rm tr}B^m)$ is order of one for the
$\Lambda$-dependence.
Therefore, the potential has a series of
$\sum t_m ({\rm tr } B^m)$. Such terms appear in the c=1 string theory from the 
calculation of the tachyon correlators \cite{Mukhi1,Dijkgraaf2}.

\vskip 5mm
\section{ Discussion }
\vskip 5mm
In this article, we have considered the explicit $p$-dependence of the
intersection numbers of moduli spaces of $p$-th spin curves based on
one and two Gaussian matrix models. 

In the one-matrix case, the limit $p\to -1$ gives
a generating function of the  Euler
characteristics. In the two-matrix case, we have obtained 
a logarithmic matrix model with  polynomial corrections, which is 
related to the generating function for the tachyon correlators  \cite{Mukhi1,Dijkgraaf2}.

 The duality,  on which the present analysis relies, is the relation between
the characteristic polynomials of two different Gaussian matrices.
The characteristic polynomials are computed as  determinants,
expressed  in terms of Grassmann variables $\psi_i^\alpha$,($i=1,...,N,\alpha=1,...k$).
In the large N limit
the $p$-th singularity is tuned through an appropriate choice of the eigenvalues of 
an external source matrix  $A$. 
One parameter remains, namely
the number of different $\lambda_\alpha$, $\alpha = 1\cdots k. $ The Fourier
transform with respect to the $\lambda_\alpha$ yields the correlation function
$U(s_1,...,s_n)$. The symmetry
between $N$ and $k$ becomes then implicit. Although this duality might be related 
to the  open/closed string duality \cite{Gaiotto,Maldacena,Hashimoto},  
we have not been able yet to reach  a clear picuture
in this direction.

\vskip 10mm
{\bf Acknowledgement}
\vskip 2mm
We thank E. Witten and D. Gaiotto for a discussion on the duality formula. We also thank
R. Penner for a discussion of the two matrix-model. S.H. is supported by Grant-in-Aid for
Scientific Research (C) of JSPS.

\vskip 10 mm
{\bf Appendix A: Three point correlation function $U(s_1,s_2,s_3)$}
\vskip 5mm
For the  three-point correlation function $U(s_1,s_2,s_3)$, we address ourselves to
the determinant terms in (\ref{U(sn)}) similar to the  two-point case.
The longest cycle in the determinant of a $3\times 3$ matrix  is
\be\label{cycle}
{\rm det}(a_{ij})|_{longest}= a_{12}a_{23}a_{31} + a_{13}a_{21}a_{32}
\ee
where $a_{ij} = \frac{1}{u_i-u_j+\frac{1}{2}(s_i+s_j)}$.

We consider the first cycle of (\ref{cycle}), ( the second cycle is almost the same),
\ba
&&\frac{1}{u_1-u_2+\frac{1}{2}(s_1+s_2)}\frac{1}{u_2-u_3+\frac{1}{2}(s_2+s_3)}
\frac{1}{u_3-u_1+
\frac{1}{2}(s_3+s_1)}\nonumber\\
&&=\frac{2}{s_1+s_2+s_3}
\int_0^\infty dx \int_0^\infty dy {\rm sh}(\frac{x}{2}(s_1+s_2+s_3))\nonumber\\
&&\times[ e^{-\frac{s_2}{2}x - \frac{s_1+s_2}{2}y - (x+y)u_1 + y u_2 + x u_3}
+ e^{-\frac{s_2}{2} x - \frac{s_2+ s_3}{2} y - x u_1 - y u_2 + (x+y) u_3}]
\ea
We express the two terms as
\be
U(s_1,s_2,s_3)= U^{I}+U^{II}.
\ee
After the shift $s_i\to \frac{s_i}{N}$,  
using the notation $N'= N (\frac{p-1}{pc})^{\frac{1}{p}}$, we have

\ba
&&U^{I}= \frac{2N'}{s_1+s_2+s_3}(\frac{1}{s_3})^{\frac{1}{p}}\int_0^\infty
dx
\int_0^\infty dy {\rm sh}(\frac{x}{2N'}s_1^{\frac{1}{p}}(s_1+s_2+s_3))
\nonumber\\
&&e^{-\frac{s_2}{2 N'} s_1^{\frac{1}{p}} x - \frac{s_1+s_2}{2N'} s_2^{\frac{1}{p}}y
- i v_1 (x+ (\frac{s_2}{s_1})^{\frac{1}{p}}y)+ i y v_2 + i (\frac{s_1}{s_3})^{\frac{1}{p}}
x v_3}
G(v_1)G(v_2)G(v_3)
\ea

\ba\label{UII}
&&U^{II}= \frac{2N'}{s_1+s_2+s_3}(\frac{1}{s_3})^{\frac{1}{p}}\int_0^\infty
dx
\int_0^\infty dy {\rm sh}(\frac{x}{2N'}s_1^{\frac{1}{p}}(s_1+s_2+s_3))
\nonumber\\
&&e^{-\frac{s_2}{2 N'} s_1^{\frac{1}{p}} x - \frac{s_2+s_3}{2N'} s_2^{\frac{1}{p}}y
- i v_1 x- i y v_2 + i ((\frac{s_1}{s_3})^{\frac{1}{p}}x + 
(\frac{s_2}{s_3})^{\frac{1}{p}} y) v_3}
G(v_1)G(v_2)G(v_3)
\ea
where $G(v_i)$ is defined by (\ref{G(v)}). Expanding $G(v_i)$ in powers of $\frac{1}{N'}$, $U(s_1,s_2,s_3)$
is expressed in terms of  the function $\phi_p(x)$.

The intersection numbers $<\tau_{n_1,m_1}\tau_{n_2,m_2}\tau_{n_3,m_3}>$ is obtained from the coefficients
of $s_1^{n_1+ \frac{m_1+1}{p}}s_2^{n_2+{m_2+1}{p}}s_3^{n_3+\frac{m_3+1}{p}}$.

In this three point correlation function,  non-trivial genus zero terms appear.
From $U^{II}$ in (\ref{UII}), we obtain the term $s_1^{\frac{1}{p}}s_2^{\frac{1}{p}}s_3^{1-\frac{1}{p}}$ 
in the large $N'$ limit. This leads to
\be
<\tau_{0,0}\tau_{0,0}\tau_{0,p-2}>_{g=0} = 1
\ee
Since there is terms of $(\frac{s_2}{s_3})^{\frac{1}{p}} y$ and $(\frac{s_1}{s_3})^{\frac{1}{p}}x$
in (\ref{UII}), these terms contribute in the large N limit as
$
s_1^{\frac{1+q_1}{p}}s_2^{\frac{1+q_2}{p}}s_3^{1-\frac{1+q_1+q_2}{p}}$, and
we obtain
the intersection numbers,
\be
< \tau_{0,q_1}\tau_{0,q_2}\tau_{0,p-2-q_1-q_2}>_{g=0} = 1
\ee
This is related to the  property of ring correlators found  in  \cite{Witten1}
\be\label{3point}
< \tau_{0,q_1}\tau_{0,q_2}\tau_{0,q_3}>_{g=0} = \delta_{q_1+q_2+q_3,p-2}
\ee
which is important for the chiral ring theory and superconformal theory for the primary fields.
From this result, the generating function $F$ is obtained as
\be
F = \sum <\tau_{0,q_1}\tau_{0,q_2}\tau_{0,q_3}>_{g=0} t_{0,q_1}t_{0,q_2}t_{0,q_3} +O(t^4)
\ee
and the superpotential $W$ can be constructed from the structure constants $C_{ijk}$ defined by
\be\label{structure}
C_{ijk} = \frac{\partial^3 F}{\partial t_i \partial t_j \partial t_k}
\ee
where we put
\be
t_i = t_{0,i-1}, \hskip 5mm (i=1,...,p-1)
\ee

If we consider only primary field, neglecting gravitational descendants, we only need the terms
$\prod_m t_{0,m}$. When we consider this primary field, in the genus zero case, we obtain therefore
,for instance for p=5,
\be
F = \frac{1}{2} t_{0,0}^2 t_{0,3} + t_{0,0}t_{0,1}t_{0,2} + \frac{1}{3!}t_{0,1}^3 + O(t^4)
\ee
and the structure constants become
$C_{114}=C_{123}=C_{222}=1$ for p=5 case.

\vskip 5mm
{\bf Appendix B: The n-point correlation function for $n\ge 4$}
\vskip 5mm

The calculation of the  n-point correlation function $U(s_1,...,s_n)$ at edge singularities follows the same steps
as for the  n=2 and 3 cases. For the discussion of the higher chiral ring structure , we need more than three points, and thus we consider  n$\ge$4.

One of the longest cycle terms in the determinant for the four point correlation function $U(s_1,s_2,s_3,s_4)$  is
\ba
&&a_{12}a_{23}a_{34}a_{41} = \frac{1}{s_1+s_2+s_3+s_4} ( a_{12} + a_{23})(a_{34}+a_{41})\nonumber\\
&& \times ( \frac{1}{u_1-u_3+\frac{1}{2}(s_1+2 s_2 + s_3)}- \frac{1}{u_1-u_3 -\frac{1}{2}(s_1+ 2 s_4 + s_3)})
\ea
with
\be
a_{ij} = \frac{1}{u_i-u_j+\frac{1}{2}(s_i+s_j)}
\ee
This term can be expressed by the integrals,
\ba
&&a_{12}a_{23}a_{34}a_{41}= -\frac{2}{s_1+s_2+s_3+s_4}\int_0^\infty dx dy dz
e^{-x(u_1-u_3)-\frac{1}{2}(s_2 -s_4)x }\nonumber\\
&&\times {\rm sinh} (\frac{1}{2}x (s_1+s_2+s_3+s_4))\nonumber\\
&& \times[ {\rm exp}( -\frac{1}{2}y (s_1+s_2)-\frac{1}{2}z (s_3+s_4)- u_1 y + y u_2 - z u_3 + z u_4)\nonumber\\
&& + {\rm exp}( -\frac{1}{2}y (s_1+s_2)-\frac{1}{2}z (s_1+s_4)- u_1 y + y u_2 - z u_4 + z u_1)\nonumber\\
&&  + {\rm exp}( -\frac{1}{2}y (s_2+s_3)-\frac{1}{2}z (s_3+s_4)- u_2 y + y u_3 - z u_3 + z u_4)\nonumber\\
&&   + {\rm exp}( -\frac{1}{2}y (s_2+s_3)-\frac{1}{2}z (s_1+s_4)- u_2 y + y u_3 - z u_4 + z u_1)]
\ea
Using the same change of variables and scalings as before, we obtain
\be
U(s_1,s_2,s_3,s_4) = U^{I}+U^{II}+U^{III}+U^{IV}
\ee
These four terms are given by $\sigma= s_1+s_2+s_3+s_4$,
\ba
&&U^{I} = -\frac{2N'^3}{\sigma}(\frac{1}{s_4})^{\frac{1}{p}}\int \frac{dv_i}{(2\pi)^4}
{\rm sinh}(\frac{x}{2N'}s_1^{\frac{1}{p}}\sigma)\prod_{i=1}^4 G(v_i)\nonumber\\
&&\times {\rm exp}[-\frac{1}{2N'}(s_2-s_4) s_1^{\frac{1}{p}} x - \frac{1}{2N'}(s_1+s_2)s_2^{\frac{1}{p}}y
-\frac{1}{2N'}(s_3+s_4)s_3^{\frac{1}{p}} z\nonumber\\
&&- i x v_1 - i (\frac{s_2}{s_1})^{\frac{1}{p}} y v_1 + i y v_2 + i (\frac{s_1}{s_3})^{\frac{1}{p}} x v_3
- i z v_3 + i (\frac{s_3}{s_4})^{\frac{1}{p}} z v_4)
\ea

\ba
&&U^{II} = -\frac{2N'^3}{\sigma}(\frac{1}{s_4})^{\frac{1}{p}}\int \frac{dv_i}{(2\pi)^4}
{\rm sinh}(\frac{x}{2N'}s_1^{\frac{1}{p}}\sigma)\prod_{i=1}^4 G(v_i)\nonumber\\
&&\times {\rm exp}[-\frac{1}{2N'}(s_2-s_4) s_1^{\frac{1}{p}} x - \frac{1}{2N'}(s_1+s_2)s_2^{\frac{1}{p}}y
-\frac{1}{2N'}(s_1+s_4)s_3^{\frac{1}{p}} z\nonumber\\
&&- i x v_1 - i (\frac{s_2}{s_1})^{\frac{1}{p}} y v_1 + i (\frac{s_3}{s_1})^{\frac{1}{p}} z v_1 + 
i y v_2 + i (\frac{s_1}{s_3})^{\frac{1}{p}} x v_3
- i (\frac{s_3}{s_4})^{\frac{1}{p}} z v_4)\nonumber\\
\ea
\ba
&&U^{III} = -\frac{2N'^3}{\sigma}(\frac{1}{s_4})^{\frac{1}{p}}\int \frac{dv_i}{(2\pi)^4}
{\rm sinh}(\frac{x}{2N'}s_1^{\frac{1}{p}}\sigma)\prod_{i=1}^4 G(v_i)\nonumber\\
&&\times {\rm exp}[-\frac{1}{2N'}(s_2-s_4) s_1^{\frac{1}{p}} x - \frac{1}{2N'}(s_2+s_3)s_2^{\frac{1}{p}}y
-\frac{1}{2N'}(s_3+s_4)s_3^{\frac{1}{p}} z\nonumber\\
&&- i x v_1 - i y v_2 + i (\frac{s_1}{s_3})^{\frac{1}{p}} v_3 x - i z v_3 + i (\frac{s_2}{s_3})^{\frac{1}{p}} y v_3
 + i (\frac{s_3}{s_4})^{\frac{1}{p}} z v_4)
\ea
\ba
&&U^{IV} = -\frac{2N'^3}{\sigma}(\frac{1}{s_4})^{\frac{1}{p}}\int \frac{dv_i}{(2\pi)^4}
{\rm sinh}(\frac{x}{2N'}s_1^{\frac{1}{p}}\sigma)\prod_{i=1}^4 G(v_i)\nonumber\\
&&\times {\rm exp}[-\frac{1}{2N'}(s_2-s_4) s_1^{\frac{1}{p}} x - \frac{1}{2N'}(s_2+s_3)s_2^{\frac{1}{p}}y
-\frac{1}{2N'}(s_1+s_4)s_3^{\frac{1}{p}} z\nonumber\\
&&- i x v_1 + i (\frac{s_3}{s_1})^{\frac{1}{p}} z v_1 - i y v_2 + i (\frac{s_1}{s_3})^{\frac{1}{p}} x v_3
+i (\frac{s_2}{s_3})^{\frac{1}{p}} y v_3 - i (\frac{s_3}{s_4})^{\frac{1}{p}} z v_4)\nonumber\\
\ea
From $U^{III}$, we obtain
the term  $s_1^{\frac{2}{p}}s_2^{\frac{2}{p}}s_3^{1-\frac{1}{p}}s_4^{1-\frac{1}{p}}$ which gives
the intersection number $<\tau_{0,1}\tau_{0,1}\tau_{0,p-2}\tau_{0,p-2}>_{g=0}$.
In this large N' limit, we have
\ba
&&U^{III} =- (\frac{s_1}{s_4})^{\frac{1}{p}}\int_0^\infty dx dy dz \int \frac{d v_i}{(2\pi)^4}
x \cdot (\frac{s_4}{2} s_1^{\frac{1}{p}} x)
\cdot (\frac{1}{2}s_3 s_2^{\frac{1}{p}} y) \cdot i (\frac{s_2}{s_3})^{\frac{1}{p}} y v_3\nonumber\\
&& {\rm exp}[ - \frac{i^p}{p}\sum_i v_i^p - i x v_1 + i (\frac{s_3}{s_1})^{\frac{1}{p}} z v_1 - i y v_2 ] \times \nonumber \\
&& {\rm exp}[  i (\frac{s_1}{s_3}) x v_3 + i (\frac{s_2}{s_3})^{\frac{1}{p}} y v_3 - i (\frac{s_3}{s_4})^{\frac{1}{p}}
z v_4 ]\nonumber\\
\ea
Expanding the factors ${\rm exp}[i (\frac{s_1}{s_3}) x v_3 + i (\frac{s_2}{s_3})^{\frac{1}{p}} y v_3 - i (\frac{s_3}{s_4})^{\frac{1}{p}}
z v_4 ]$ we obtain the series of the intersection numbers for the primary fields
in the genus zero case,
\be\label{4tau}
<\tau_{0,q_1}\tau_{0,q_2}\tau_{0,p-q_1-q_2+q_3},\tau_{0,p-2-q_3}>_{g=0}
\ee
where $q_1,q_2=1,2,...$, and $q_3=0,1,2,...$.
The other three terms $U^{I},U^{II}$ and $U^{IV}$ do not yeild terms of the type $s_1^{\frac{q_1+1}{p}}s_2^{\frac{q_2+1}{p}}
s_3^{1-\frac{q_1+q_2-q_3}{p}}s_4^{1-\frac{q_3+1}{p}}$.

For p=2, the term (\ref{4tau}) does not exist, since $\tau_{0,1}$ is not allowed. For higher n-point correlations
($n\ge 5$), there is no correction for the same reason. 
Therefore, for p=2, the function $F$ for the
primary field is
\be
F = \frac{1}{6} t_{0,0}^3 \hskip 5mm (p=2).
\ee
For p=3, we obtain from (\ref{4tau}) and (\ref{3point}),
\be
F= \frac{1}{2} t_{0,0}^2 t_{0,1} + \frac{1}{72} t_{0,1}^4
\ee
For p=4, we obtain (\ref{4tau}) and (\ref{3point}),
\be\label{F40}
F= \frac{1}{2}t_{0,0}^2 t_{0,2} + \frac{1}{2}t_{0,0}t_{0,1}^2+ \frac{1}{16}
t_{0,1}^2 t_{0,2}^2 + \frac{1}{8\cdot 5!} t_{0,2}^5
\ee
The last term is evaluated from the five point correlation function, which has the form, for general $p$,
\be
s_1^{1-\frac{1+q_1}{p}}s_2^{\frac{3+q_2}{p}}s_3^{1-\frac{q_2+1}{p}}s_4^{\frac{3+q_1}{p}}s_5^{1-\frac{1}{p}}
\sim t_{0,p-2-q_1}t_{0,2+q_2}t_{0,p-2-q_2}t_{0,2+q_1}t_{0,p-2}
\ee
which leads to the last term in (\ref{F40}) for p=4.
We have investigated the intersection numbers of primary fields, but other gravity descendants can be
obtained in the same ways, which would be in factor  of $t_{n,m}$ ($n\ne 0$).
\vskip 5mm
{\bf Appendix C: Ginzburg-Landau potential for primary fields and Gelfand-Dikii equation}
\vskip 5mm
The structure constant $C_{ijk}$ defined by (\ref{structure}) are obtained from the n-point correlation function
through the intersection numbers with n marked points.
In this appendix, we discuss the relation to the superpotential \cite{Witten1}.
Using the notation
\be
t_i = t_{0,i-1} \hskip 5mm (i=1,2,...)
\ee
and the metric $\eta^{nm}=\delta_{n+m,p}$,
we define
\be
C_{ij}^k = \sum_{m=1}^{p-1} C_{ijm}\eta^{mk}.
\ee
In this notation, F becomes, in the p=4 case for instance,
\be
F= \frac{1}{2}t_1^2 t_3 + \frac{1}{2} t_1 t_2^2 + \frac{1}{4}t_2^2 t_3^2 + \frac{1}{60} t_3^5
\ee

We find that the Witten, Dijkgraaf, Verlinde,@Verlinde  relation@\cite{Witten1,Dijkgraaf1}
\be
C_{ij}^m C_{mkl} = C_{ik}^m C_{mjl}
\ee
holds for the structure constants that we have computed.

Then the $C_{ij}^k$ have a ring structure,
\be\label{ring}
\phi_i\phi_j = \sum_k C_{ij}^k \phi_k \hskip 5mm (mod [W'(x)])
\ee
where $\phi_i$ is defined by the derivative of the Landau-Ginzburg potential $W(x)$ 
\be
\phi_i = -\frac{\partial W}{\partial t_i}
\ee

We have obtained the function $F$ by the evaluation of the intersection 
numbers of primary fields up to the
6-point correlation function.
For the  p=5 case,
\ba
&&F= \frac{1}{2}t_{0,0}^2 t_{0,3}+ t_{0,0}t_{0,1}t_{0,2} + \frac{1}{6}t_{0,1}^3 + \frac{1}{4}t_{0,1}^2
t_{0,3}^2 \nonumber\\
&&+\frac{1}{2}t_{0,1}t_{0,2}^2 t_{0,3}+\frac{1}{6}t_{0,1}t_{0,2}^3 + \frac{1}{2}t_{0,1}^2 t_{0,2}t_{0,3}
\nonumber\\
&&+ \frac{1}{12}t_{0,2}^4 + \frac{1}{6}t_{0,2}^2t_{0,3}^3 + \frac{1}{120}t_{0,3}^6
\ea
This leads to $C_{213}=C_{411}=C_{222}=1,C_{224}=t_4,C_{231}=t_3,C_{232}=t_4,C_{244}=t_2,C_{334}=t_2+t_4^2,
C_{333}= 2 t_3,C_{332}= t_4, C_{341}=t_3 t_4,C_{342}= t_3,C_{444}= t_3^2+t_4^3,C_{434}= 2 t_3 t_4$.

The ring structure (\ref{ring}) holds with
\be\label{Wp5}
W(x) = \frac{1}{5} x^5 - t_4 x^3 - t_3 x^2 + (t_4^2-t_2) x + (t_3 t_4 - t_1)
\ee
The function $\phi_i$ is
\be
\phi_i= - \frac{\partial W}{\partial t_i}
\ee
and the equation of the ring structure (\ref{ring}) holds with $ {\rm mod} W'(x) = {\rm mod}[ x^4-3 t_3 x^2 - 2 t_3 + t_4 - t_2]$.

The Landau-Ginzburg potential $W$ in (\ref{Wp5}) is the same as the superpotential of
the twisted N=2 superconformal theory
for $A_{4}$ type. From the singularity theory, this potential (\ref{Wp5}) is
called a swallow tail.

Thus we find that the random matrix theory with an external source 
for the p-th critical point gives the 
Landau-Ginzburg potential of the $N=2$ superconformal theory for the primary 
fields in the genus zero case.
Our integral expression for the  n-point correlation function may be used without difficulties  
 to give the intersection numbers and
 the gravity descendants  
for  higher genus.

We note that these algebraic structures reduce to the  Gelfand-Dikii equation, which
gives  the generalized KdV hierarchies. For instance, in the case p=3, we obtain from
our formulation the
Boussinesque equation,
\be
\frac{\partial^2 F}{\partial t_{0,1}^2} = \frac{\partial^4 F}{\partial t_{0,0}^4} - 
\frac{2}{3}(\frac{\partial^2 F}{\partial t_{0,0}^2})^2
\ee
and it's higher gravitational desendents. This hierarchy can be derived from
Gelfand-Dikii equation \cite{Witten1}. This equation is expressed by \cite{Gelfand}
\be
i \frac{\partial Q}{\partial t_{n,m}}= [ Q_{+}^{n+ \frac{m+1}{p}},Q] \cdot \frac{C_{n,m}}{
\sqrt{p}}
\ee
which is the generalization of Lax equation. The Q is given
\be
Q = D^p - \sum_{i=0}^{p-2}u_i(x) D^i
\ee
and
the fraction power of $Q$ is
\be
Q^{\frac{1}{p}} = D + \sum_{i>0}w_i D^{-i}
\ee
From this formulation, we obtain the relation to $F$ as
\be
\frac{\partial^2 F}{\partial t_{0,0}t_{n,m}} = - C_{n,m}{\rm res}(Q^{n+\frac{m+1}{p}})
\ee
where
\be
C_{n,m} = \frac{(-1)^n p^{n+1}}{(m+1)(p+m+1)\cdots (pn+ m+ 1)}
\ee

\end{document}